# Vortex Interaction with Mesoscopic Surface Irregularities in High Temperature Superconductors

L.N.Shehata[(*)] and A.Y. Afram

Department of Mathematics and Theoretical Physics

Atomic Energy Authority, Egypt

E-Mail: louis_shehata@yahoo.co.uk

**Abstract**

The conformal mapping method is used to study the problem of flux line interaction with surface cavities having cylindrical profile and characteristic size $\ell \ll \lambda$, i.e, within mesoscopic scale, and $\lambda$ is the penetration length. It is shown that the metastable states are achieved when the dimensions of the surface irregularities do not exceed the coherence length $\xi$. Our study shows that the surface barrier may vanished at some weak point at which the surface irregularities have mesoscopic scales. On the other hand, the surface barrier is completely disappeared when the surface defects size $\ell \gg \lambda$. Our results are compared with the available experimental data and theoretical results.

## 1. Introduction

Experimental measurements of the magnetization curves, even for ideally smooth surfaces in high temperature superconductors (HTSs), show remarkable hysteresis [1-3].

Early, hysteresis was observed in ideal type II low temperature superconductors (LTSs) [4-6]. This phenomenon is usually explained [7-10] by the existence of the Bean-Livingston (BL) surface potential barrier at the surface of the superconductors (SCs). The BL surface barrier (SB) prevents vortices (or flux lines (FLs)) to enter into or exit from the SC. Therefore, the SC is found to be in a metastable state with magnetic induction values different from those corresponding to the macroscopic equilibrium state at given value of the external field[6]. As it was shown by Bean and Livingston[7] and De Gennes [8], the BL barrier arises from the competition between the repulsion of a FL from the surface due to its interaction with the decaying external field and the attraction to the surface due to the interaction of the FL with its antivortex mirror image. The maximum value of the SB takes place at a distance $\xi$ from the planar surface, where $\xi$ is the coherence length. Although the flux penetration becomes thermodynamically favorable at $H_{c1}$, where $H_{c1}$ is the first critical field, the penetration starts, instead, at the superheating field $H_p > H_{c1}$ at which the SB disappears. The exact value of $H_p$ was calculated by Galiako [11] for ideal type II LTSs with large values of Ginzburg-Landau parameter $\kappa$. It was found that $H_p = 0.8\ H_c$ at $T = 0$ and $H_p = 0.745\ H_c$ at $T_c - T \ll T_c$, where $T_c$ is the critical temperature and $H_c$ is the thermodynamic equilibrium magnetic field. On the other hand, for non-ideal SCs, the surface roughness must suppress the BL barrier and therefore, decreases the value of the penetration magnetic field $H_p$[12]. This assumption has been confirmed by many experiments in which the SC surface has been damaged or contains roughness and irregularities [13-18]. However, experimental studies yielded $H_p$ of smaller values, may be in many times, than those theoretically predicted (see for example ref.(15)). The reason of these discrepancies, in addition to other effects, is the use of SCs with surface irregularities of small scale distortion. The effect of the surface rouphness on the BL surface barrier [20] and the FL entry conditions in type II SCs [21] have been theoretically studied for large scale surface distortions.

It is interesting, therefore, to explain how the value of the potential SB is affected by the curvature of the surface. The curvature scale may be large or small as its characteristic dimension is compared with the penetration depth $\lambda$, and the coherence length $\xi$. We have two different cases. In the first case, the surface may contain large-scale cavities (or distortions) with characteristic dimension $\ell \gg \lambda$. In this case the problem may be considered in the usual way where the value of external magnetic field $H_0$ reaches the value $H_p$, at which the SB disappears. In this problem, the standard methods are used and it will not be considered here. We only notice, that in such cases it is possible to attain penetration fields $< H_p$.

The second case concerns with the other limit, when $\ell \ll \lambda$ (curvatures with mesoscopic dimensions), and the value of the external magnetic field is the same at all the surface points of the SC. However, such surface distortions may yield to significant different surface current distribution and, subsequently, to different value of the potential SB. This situation appears, for example, when considering the case, in which the surface has sharp curved profile and $H_0$ is parallel to the generator of the cylindrical surface of the SC.

In the present work, we will study the interaction between a straight FL, parallel to the direction of external field $H_0$ and a cylindrical cavity at the surface of the SC. The paper is organized as follows. The next section (section 2) deals with the basic model and the approximations that will be used to study selected cases. In section 3 the interaction with cylindrical cavities will be considered by using the conformal mapping method. Detailed procedure will be used to construct the conformal transformation which will be used to solve the approximated formulae that were obtained in section 2. The free energy and the force acting on the FL will be evaluated in section 4 and the calculation of the Gibbs free energy is demonstrated in section 5. In the last section some concluding remarks are presented.

## 2. The Model

Consider the magnetic field **H** to be directed along the z-axis of the superconductor and the quantity $\varphi_0 / 2\pi\lambda^2$ is used to rescale the magnetic field, i.e,

$$\mathbf{H} = \frac{\varphi_0}{2\pi\lambda^2}\mathbf{h}, \quad \mathbf{h} = \operatorname{curl}\mathbf{a}, \quad \mathbf{j} = \operatorname{curl}\mathbf{h} \tag{1}$$

Where **a** is the vector potential, **h** the internal field and **j** the current density, therefore the London equation can be written as:

$$\operatorname{curl}\mathbf{h} = \nabla\phi - \lambda^{-2}\mathbf{a} \tag{2}$$

Where ϕ is the phase of the pairing potential, which changes by $2\pi$ as it moves around the center of the vortex line [19]. It is easy to show that at small distances $\rho \ll \lambda$ the first term in the right hand side (rhs) of eq. (2), $|\nabla \phi|$, is proportional to $\rho^{-1} \gg |\mathbf{a}|\lambda^{-2}$, and consequently could be neglected. Therefore, we get the following formulae for the current components, i.e.

$$j_x = \frac{\partial \phi}{\partial x} = \frac{\partial h}{\partial y}, \qquad j_y = \frac{\partial \phi}{\partial y} = \frac{\partial h}{\partial x} \qquad (3)$$

According to eq.(3) we may introduce the analytic complex function,

$$\Omega(z) = \phi + i\,h \qquad \text{and} \qquad \frac{\partial \Omega}{\partial z} = j_y - i\,j_x \qquad (4)$$

Where $z = x + i\,y$. In particular, if we consider a straight FL at $z_0 = x_0 + i\,y_0$ in a boundless SC, we get [8,15],

$$\Omega(z) = \Omega_0(z - z_0) = -i \ln \frac{\gamma(z - z_0)}{2\lambda} \qquad (5)$$

Where, $\ln \gamma = 0.577$ is the Euler's constant and $(x_0, y_0)$ are the coordinates of the centre of the FL. The above formula (eq.(5)) is valid only if $\xi = \frac{\lambda}{\kappa} \ll |z - z_0| \ll \lambda$.

### 3. Interaction with Cylindrical Cavity

Consider a boundless SC in which there is a cylindrical cavity with its generator directed parallel to the z-axis. Denote by D the region which describes this cavity in the xy-plane and by Γ - the boundary (or the profile) of the cavity (see Fig.(1)). Consider first the case in which the dimension of the cavity is smaller than λ and the FL is localized at distance from the boundary smaller than λ. Under these assumptions, the approximated formulae given by eq.(3) may be applied. Therefore, the analytical complex function $\Omega(z)$ defined by eq.(4) should satisfy the following conditions:

1. The function $\Omega(z)$ should be a multivalued analytical function outside the region D. To separate the single value branch of $\Omega(z)$, we make the cut S, which joins the centre of the FL at $z_0$ with its finite end point (Fig.1). In this case, the difference between the values of $\Omega(z)$ at these two terminal points is $2\pi$.

2. The normal component of the current, $J_n$, should vanish at the boundary Γ, i.e.

$$J_n = \frac{\partial \Phi}{\partial n} = 0 \qquad (6)$$

Using eqs. (3) and (4) one may rewrite the last condition in the form,

$$\mathrm{Im}\,\Omega(z)\big|_\Gamma = h_i = \mathrm{const.} \qquad (7)$$

where $h_i$ is the value of the dimensionless magnetic field inside the cavity.

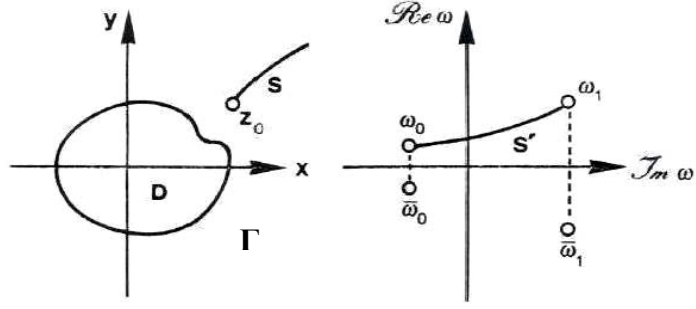

Fig. 1: Conformal transformation from the x,y- plane into the ω plane

Furthermore, and in order to determine the function $\Omega(z)$, consider the conformal transformation $\omega = \omega(z)$, which transform the exterior of the region D into the ω-upper half plane. In this case, the infinite points of the z-plane will be transformed into the point $\omega_1 \equiv \omega(\infty)$ and the point $z_0$ into $\omega_0 \equiv \omega(z_0)$. Therefore, the cut S transforms into the finite curve S', which lies totally in the ω-upper half-plane. Note that the boundary Γ of the cavity in xy-plane is transformed into the Im ω-axis in the ω-plane. (Fig.(1))

In addition to the above arguments the following two conditions will be introduced:

$$\left.\frac{d\omega}{dz}\right|_{z=z_0} = \omega'(z_0) \neq 0 \quad \text{and} \quad \left.\frac{d\omega}{d(1/z)}\right|_{z\to\infty} = \omega_1' \neq 0 \tag{8}$$

Now, it is easy to show that the following form of $\Omega(z)$ satisfies all the above mentioned conditions,

$$\Omega(z) = i\ln\left|\frac{(\omega-\bar{\omega}_0)(\omega-\omega_1)}{(\omega-\omega_0)(\omega-\bar{\omega}_0)}\right| - i\ln\left|\frac{\gamma(\omega_1-\bar{\omega}_0)\omega_1'}{2\lambda(\omega_1-\omega_0)(\omega_1-\bar{\omega}_1)}\right| \tag{9}$$

Therefore, the magnetic field inside the cavity, $h_i$, is given by the second term in the rhs of eq.(9) by using that $\omega_1 = \bar{\omega}_1$ and $\omega_0 = \bar{\omega}_0$ (see eq. (7)), i.e.,

$$h_i = i \ln \left| \frac{2\lambda}{\gamma} \frac{(\omega_1 - \omega_o)(\omega_1 - \overline{\omega}_1)}{\omega_1'(\omega_1 - \overline{\omega}_o)} \right| \tag{10}$$

**4. The Free Energy and the Force of Interaction**

The variation in the free energy density (per unit length of the of the FL length) of a SC that containing a FL is,

$$\delta F = \frac{\varphi_0}{8\pi} H(0) \tag{11}$$

where H(0) is the value of the magnetic field at the centre of the FL. In boundless SC, $H(0) = 2\, H_{c_1}$ and in the present case we have

$$H(0) = 2 H_{c_1} + \frac{\varphi_o}{2\pi\lambda^2} \operatorname{Im} \Omega(z_o) \tag{12}$$

where,

$$2H_{c_1} = \frac{\varphi_o}{2\pi\lambda^2} \ln \frac{2\lambda}{\gamma \xi} \tag{13}$$

is the first critical field. We may write the following formula for the free energy density F as follows,

$$F = \frac{\varphi_o}{2\pi} H_{c_1} + \delta F \tag{14}$$

where $\delta F$, by using eq.(11), reads

$$\delta F = \left(\frac{\Phi_o}{2\pi\lambda}\right) \ln\left|\frac{(\omega_o - \overline{\omega_o})(\omega_o - \omega_1)^2(\omega_1 - \overline{\omega_1})}{\omega'(z_o)(\omega_o - \overline{\omega_1})\omega_1'}\right| \tag{15}$$

The force of interaction with the surface is f, where

$$\mathbf{f} = -\nabla_o(\delta F) \tag{16}$$

where

$$\nabla_o = \hat{\mathbf{i}}\frac{\partial}{\partial x_o} + \hat{\mathbf{j}}\frac{\partial}{\partial y_o} \tag{17}$$

To illustrate the above formulae, consider first, the case in which the profile of the cavity is the circle $|z| = R \ll \lambda$, and $R < |z_0| = r \ll \lambda$. Then, if we take

$$\omega(z) = i\frac{z-R}{z+R} \tag{18}$$

and using eqs. (10 -12) we get (compare with ref. (21)),

$$\delta F = \left(\frac{\varphi_o^2}{4\pi\lambda}\right)^2 \ln\left(1 - \frac{R^2}{r^2}\right) \tag{19}$$

and,

$$H_i = \frac{\varphi_0}{2\pi\lambda} \ln\frac{2\lambda}{\gamma r} \tag{20}$$

Secondly, consider the case in which the boundary of the cavity is an arbitrary smooth curve. Let ρ denotes the distance between the centre of FL, $z_0$, and the nearest point $z_0'$ from the boundary and let the radius of the cavity at this point be R. If $\rho \ll R(z_0')$, we get,

$$|\omega_o - \overline{\omega_o}| \simeq 2\rho\,|\omega'(z_o)| \tag{21}$$

Therefore,

$$\delta F = \left(\frac{\varphi_o}{2\pi\lambda}\right)^2 \ln \frac{2\rho(\omega_1 - \bar{\omega}_1)}{|\omega_1'|} \qquad (22)$$

and,

$$f = -\frac{\varphi_o}{4\pi\lambda}\frac{1}{\rho} \qquad (23)$$

The last expression, which is valid only for $\xi \ll \rho \ll R \ll \lambda$, shows that the force of interaction with the cavity which has small dimension coincides with the force acting on the FL which is localized near the SC boundary[8].

The third case deals with large scale cavity, i.e., its characteristic dimension are greater than $\lambda$ (there is no flux creep). In this case we may put $h_i = 0$, and assume that the conformal transformation $|\omega_1| \gg |\omega_0|$ and the second condition in eq.(8) is not applicable. Therefore, eq.(9) may be replaced by the following transformation,

$$\Omega(z) = i \ln \left|\frac{\omega - \bar{\omega}_o}{\omega - \omega_o}\right| \qquad (24)$$

Consequently,

$$F = \left(\frac{\varphi_o}{4\pi\lambda}\right)^2 \ln \left|\frac{\omega_o - \bar{\omega}_o}{\xi\omega'(z_o)}\right| \qquad (25)$$

Now, if the boundary of the cavity is smooth, then by using eq.(21) the force of interaction will have the same form given by eq.(23). On the other hand, if the boundary is not smooth and has an edge shape, then near the top of the edge we may have,

$$\omega(z) \sim z^\nu; \qquad \tfrac{1}{2} < \nu < 1 \tag{26}$$

And from eq.(25) we get,

$$F = \left(\frac{\varphi_o}{4\pi\lambda}\right)^2 \ln\left|\frac{2\rho\sin\nu\varphi}{\xi\nu}\right| \tag{27}$$

Where $z_0 = \rho e^{i\varphi}$. In this case the force of interaction has a component parallel to the boundary.

### 5. The Gibbs Potential

If the flux line is localized near the exterior boundary of the SC which is in external field $H_0$ directed parallel to the surface, then it is necessary to use the Gibbs free energy,

$$G = F + \frac{\varphi_o}{4\pi}(H_L + H_o) \tag{28}$$

Here $H_L$ is the London penetrating field. Using the dimensionless units given by eq. (1) we can rewrite the last equation in the form,

$$G = F + \frac{\varphi_o}{8\pi^2\lambda^2}(h_L + h_o) \tag{29}$$

Where $h_L$ is the dimensionless London's field, which satisfies the following equation and its boundary conditions

$$\Delta h_L - \lambda^{-2} h_L = 0; \quad h_L|_\Gamma = h_o; \quad h_L|_{|z|\to\infty} = 0 \tag{30}$$

If the radius of curvature of the boundary profile R << λ and the distance between the FL and the surface is small (ρ << λ), then using eqs. (30) we get,

$$h_L - h_o \simeq \frac{-h_o \rho}{\lambda} \tag{31}$$

Therefore, by using eq. (21) and eq. (25) we get,

$$G = \left(\frac{\varphi_o}{4\pi\lambda}\right)^2 \left[\ln\frac{2\lambda}{\xi} - \frac{2h_o\rho}{\lambda}\right] \tag{32}$$

The last equation shows that the Gibbs potential has the same form as in the case of a plane boundary[8]. To determine the value of the external field, at which the SB vanishes, we use the condition,

$$\left.\frac{\partial G}{\partial \rho}\right|_{\rho=\xi} \leq 0 \tag{33}$$

It is known that London's approximation is not valid for ρ < ξ (ρ ≠ 0). From eq. (3) and by using eq.(32) we get,

$$h_o < \frac{\lambda}{2\xi} \tag{34}$$

Or in the usual units

$$H_o > H_p \simeq \frac{H_c}{\sqrt{2}} \tag{35}$$

Now, for sharp curved (~ λ) boundary profile with characteristic size of the distortion $\ell << \lambda$, the distribution of the current density may be obtained by using the approximation given by eqs. (3) and we will assume that

$$\Omega_L = \Phi_L + i h_L, \quad j_{Lx} - i j_{Ly} = \frac{\partial \Omega_L}{\partial z} \quad (36)$$

Where, $\Omega_L$ satisfies the following condition,

$$\mathrm{Im}\, \Omega_L \big|_\Gamma = h_o, \quad \frac{\partial \Omega_L}{\partial z}\bigg|_{|z| \to \infty} = -\frac{h_0}{\lambda} \quad (37)$$

These conditions are easily deduced from eqs. (30) and (31).

From the above, it is clear that the problem we are dealing with is the analogous of the well known hydrodynamic problem of the flux flow -being homogeneous at infinity- of an ideal liquid around a dam. Therefore, let us more exactly choose the function ω(z) by introducing an additional condition, that is

$$\frac{d\omega}{dz}\bigg|_{|z| \to \infty} = 1 \quad (38)$$

In this case, we get

$$\Omega_L = i h_o - \left(\frac{h_o}{\lambda}\right) \omega(z) \quad (39)$$

Using eqs.(25) and (29) the Gibbs potential is obtained in the form

$$G = \left(\frac{\varphi_o}{4\pi\lambda}\right)^2 \left[ \ln \left| \frac{\omega_o - \overline{\omega}_o}{\xi \omega'(z_o)} \right| - \frac{2h_o}{\lambda} \mathrm{Im}\, \omega_o \right] \quad (40)$$

The Gibbs potential in the above formula can be used for the case of a smooth boundary by using eqs. (21) and (33), we get

$$H_0 \geq H_p(z) = H_p |\omega'(z)|^{-1} \quad (41)$$

$H_p(z)$ is the value of the external field at which the surface potential barrier disappears at the point z at the surface profile. If we denote by $H_{p,min}$ the lowest value for $H_p(z)$, then, from eq.(41), it is easy to conclude that for the case of Fig.(2b) we get

$$H_{p,\,min} = \frac{bH_p}{a+b} \qquad (42)$$

and for the case of Fig.(2c) we find

$$H_{p,\,min} = \frac{bH_p}{\left(a^2+b^2\right)^{1/2}} \qquad (43)$$

The above expressions show that the minimum value for the surface potential barrier will be achieved at the bottom of the corresponding scratches. It is important to emphasize that these formulae are applicable under the condition that $R_0 \gg \xi$, where $R_0$ is the lowest value of the profile radius of curvature and $a \ll \lambda$. In both cases we have

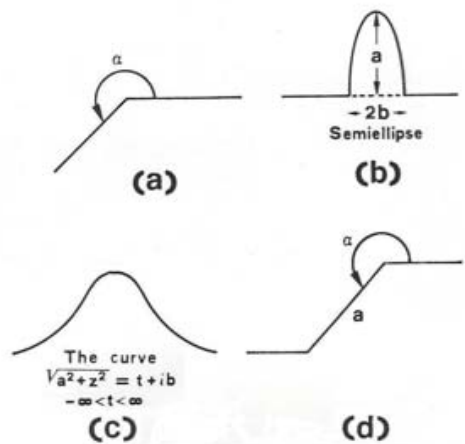

Fig. 2 : Selected surface irregularities

,

$$H_{p,\,min} \leq \frac{H_p}{\sqrt{\kappa}} \qquad (44)$$

Consider now the case of Fig.(1d). In this case the reverse transformation of $\Omega = \Omega(z)$ yields the following integral,

$$z = \int_0^\omega \left[\frac{t}{\alpha+t}\right]^{\frac{1}{\nu}-1}; \quad \alpha = \frac{a\nu \sin\frac{\pi}{\nu}}{\pi(\nu-1)} \qquad (45)$$

where, $\frac{1}{2} \leq \nu \leq 1$ and $\xi \leq a \prec \lambda$. The lowest value of the superheating field $H_{p,min}$ for this case can be obtained by using eq.(2). We notice that in this case eq.(41) is not applicable, however, using eqs. (16), (33) and (40), one obtains,

$$H_{p,\,min} = H_s \left(\frac{\alpha\nu}{\xi}\right)^{1-\nu} \qquad (46)$$

and this value of $H_{p,min}$ is achieved along the ray $\arg z = \pi/2\nu$. We notice that the inequality (40) is valid in this case.

At last, we consider the case in which the profile of the surface contains an angle (see Fig. (2a)). In this case one has to determine $h_L$ by solving eq.(30), which gives,

$$h_L(z) = \left(\frac{2h_o}{\pi}\right) \int_0^\infty \frac{\cosh\frac{\pi\tau}{2}}{\sinh\frac{\pi\tau}{2\nu}} \left[\sinh(\varphi\tau) + \sinh\left(\frac{\pi}{\nu}-\varphi\right)\tau\right] K_{i\tau}\left(\frac{\rho}{\lambda}\right) d\tau \qquad (47)$$

where $z = \rho e^{i\varphi}$, $0 \leq \varphi \leq \frac{\pi}{\nu}$, and $K_{i\tau}(x)$ is Macdonald's function of pure imaginary index. Using the above result together with eq. (27) we find,

$$H_{p,\,min} = \frac{\Gamma(\nu)\sin\left(\frac{\pi\nu}{2}\right) H_p}{\nu(2\kappa)^{1-\nu}} \qquad (48)$$

or,

$$H_{p,\min} \geq \frac{H_s}{\kappa^{1/2}} \tag{49}$$

Comparing the last inequality with eq.(46) we conclude that as the height of the step increases the quantity $H_p$ stops its decreasing starting from the value a ~ λ/2π.

## 6. Conclusions

The method of the conformal mapping was used to treat the problem of flux line interaction with surface cavities having cylindrical profile and mesoscopic characteristic size $\ell \ll \lambda$. The main conclusions of our study are summarized in the following points.

1. The metastable states that correspond to the value of the superheating magnetic field, $H_p$, are achieved only when the surface of the SC contains irregular regions of dimension not exceeding the coherence length $\xi$.

2. The presence of surface mesoscopic distortions of dimensions in the interval $(\xi, \lambda)$, make the potential surface barrier, at some "weak points", disappears at certain value of the external field, such that, $\frac{H_p}{\sqrt{\kappa}} < H_o < H_p$. This value of $H_o$ is greater than the first critical field $H_{c_1}$ in $\sqrt{\kappa}$ times.

3. The surface barrier is completely vanished at the presence of large-scaled surface distortions ( e.g., lugs and steps perpendicular to the direction of the external magnetic field and of sizes $\ell \gg \lambda$ )

4. Scratches of semi-circular profiles radius of the order of the material particle dimension my remain even the same after the sample surface is mechanically polished or after chemical and electrochemical treatment[18]. In these cases, and according to eqs.(42) and (43) we get $H_{p,min} = \frac{1}{2} H_p$. These results agree with experimental results [16-18].

# References

1. D.X. Chen, A. Hermando, F. Conde, J. Ramfrez, J. M. Gonzalez-Calbet and M. Vallet, J. Appl. Phys. 75, 1, 1994.

2. B.J. Baelus, K. Kadowaki and F. M. Peeters, Phys. Rev. B71, 024515, 2005.

3. B.J. Baelus, K. Kadowaki and F. M. Peeters, Phys. Rev. B71, 024514, 2005.

4. H.A. Ulmaier and W.F. Gauster, J. Appl. Phys. 37, 4519, 1966.

5. H.A. Ulmaier, Phys. Stat. Sol. 17, 631, 1966.

6. A. M. Campbell, I.E. Evetts and D. Dew-Hughas, Phil. Hag. 18, 313, 1968.

7. C. P. Bean and J. D. Livingston, Phys. Rev. Lett. 12, 14, 1964.

8. P.G. de Gennes, Superconductivity of Metals and Alloys, Benjamin, New York, 1966, Chapter 3.

9. F. F. Ternovski and L. N. Shehata, Sov. Phys. JETP 35, 1202, 1972.

10. J. R. Clem, Low Temp. Phys.-LT-13, Vol. 3, Eds. K. D. Timmerhaus, W. J. O'Sullivan and E. F. Hammel (Plenum Pub. Corp., New York, 1974 page 102.

11. Galiako, Ekspt. Teoret. Fiz. (USSR) 50, 717, 1966.

12. A.V. Gurevich and V. T. Kovachev, Phys. Stat. Sol. (b) 145, K47, 1988.

13. M. Konczkowski, L. Burlachkov, Y. Yeshurun and F. Holtzberg, Phys. Rev. B43, 13707, 1991.

14. M. Konzkowski, L. Burlachkov, Y. Yeshurun and F. Holtzberg, Phys. C, 194, 155, 1992.

15. Kishio and K. Kitazawa, Phys. C 185-189, 1835, 1991.

16. J. F. Bussiere, Phys. Lett., 58A, 343, 1976.

17. J. F. Bussiere and M. Suenaga, J. Appl. Phys. 47, 707, 1976.

18. J. F. Bussiere and V. T. Kavachev, J. Appl. Phys. 49, 2526, 1978.

19. A.A.Abrikosov, J. Ekspt. Teoret. Fiz. (USSR) 32, 1442, 1957 and 46, 1464, 1964.

20. F. Bass, V.D.Freilikher, B.Ya.Shapiro, and M. Shvartse Phys. C 260, 231, 1996.

21. B.Y. Vodolazov, I.L. Maksimov, and E.H.Brandt Phys. C. 384, 211, 2003.

22. G.S.Mkritchian and V.V.Schmidt, J. Ekspt. Teoret. Fiz. (USSR) 61,388,1971.